\documentclass{elsart1p}
\usepackage{graphicx}
\usepackage{amssymb}
\usepackage{amsmath}
\usepackage[american]{babel}
\usepackage{array}
\usepackage{amsbsy}
\usepackage{wrapfig}
\usepackage{subfig}
\usepackage{float}
\usepackage{units}
\usepackage{graphicx}
\begin{document}
\begin{frontmatter}
%
%
%
\title{Measurement of the Structure of the Proton at HERA}
%
%
\author{Micha\l{} W\l{}asenko for the H1 and ZEUS collaborations}
\address{University of Bonn, Nussallee 12, 53115, Bonn, Germany}
\begin{abstract}
A combination of the reduced $e^{\pm}p$ cross section, previously
measured by H1 and ZEUS, has been performed and leads to well
constrained QCD fits of the parton densities in the proton. 
These results, as well as the high-$Q^2$ neutral current deep
inelastic scattering cross sections are presented. First direct
measurements of the longitudinal structure function $F_L$ are also
reported.  
\end{abstract}
\begin{keyword}
%
HERA \sep H1 \sep ZEUS \sep $F_2$ \sep $xF_3$ \sep $F_L$ \sep PDF \sep
NC \sep
\PACS 12.38.Qk \sep 12.15.Mm \sep 13.60.Hb
\end{keyword}
\end{frontmatter}
%
\section{Introduction}
\label{sec:intro}
HERA, running from 1992 until 2007, was the only
electron-proton\footnote{Unless it is  stated otherwise, terms
  \emph{electron} and \emph{positron} are used interchangeably.}
($e^{\pm}p$) collider in the world, providing head-on collision data
for two multi-purpose detectors, H1 and ZEUS.
Studies of neutral current (NC, $e^{\pm}p~\to~e^{\pm}X$, mediated by
$\gamma$ and $Z^0$ bosons) and charged 
current (CC, $e^{\pm}p~\to~\tilde{\nu_e}(\nu_e)X$, mediated by
$W^{\pm}$ bosons) deep inelastic scattering (DIS) processes,
performed by both collaborations, allowed for a substantial increase in
the understanding of pQCD and parton density functions (PDFs). 
First analyses using the full luminosity of $\approx 1~\unit{fb}^{-1}$ are
about to be finalized. This paper points to the most recent measurements of
the $F_L$ and $xF_3$ structure functions, as well as to the first 
combination of reduced $ep$ cross sections measured separately by H1 and 
ZEUS and to common QCD PDF fits, resulting in much more precise picture
of the structure of the proton. 

\section{Deep Inelastic Scattering at HERA}
\label{sec:DIS}
The kinematics of DIS can be described using three variables: the
momentum transfer, $Q^2$, the Bjorken scaling variable,
$x$ and the inelasticity, $y$ ($Q^2~=~sxy$, where $s$ is the
center-of-mass collision energy squared). The electroweak (EW)
Born-level reduced cross section for the $e^ \pm p$ NC interaction can
be written as:   
\begin{align}
\tilde{\sigma}^{e^\pm p}_r = 
\frac{xQ^4}{2\pi\alpha^2} \frac{1}{Y_+} \frac{d^2\sigma_{NC}^\pm}{dxdQ^{2}}
=   \tilde{F_{2}}(x,Q^{2})  \mp \frac{Y_{-}}{Y_+} x\tilde{F_{3}}(x,Q^{2})  
  - \frac{y^{2}}{Y_+}\tilde {F_{L}}(x,Q^{2}) .
\end{align}
where $Y_{\pm} = 1 \pm (1 - y)^{2}$. The generalized structure functions
$\tilde{F_{2}}(x,Q^{2})$, $x\tilde{F_{3}}(x,Q^{2})$ are directly
related to quark distributions. Scaling
violation of $\tilde{F_{2}}$ at small-$x$, as well as the
longitudinal structure function $\tilde{F_{L}}(x,Q^{2})$, are directly
related to the gluon density in the proton. At high $Q^2$,
$x\tilde{F_{3}}$ provides information on valence quark densities.

\section{HERA combined NC cross sections and PDF fits}
\label{sec:NCcombined}
\begin{figure}[h]  
\centering
\includegraphics[width=0.51\textwidth]{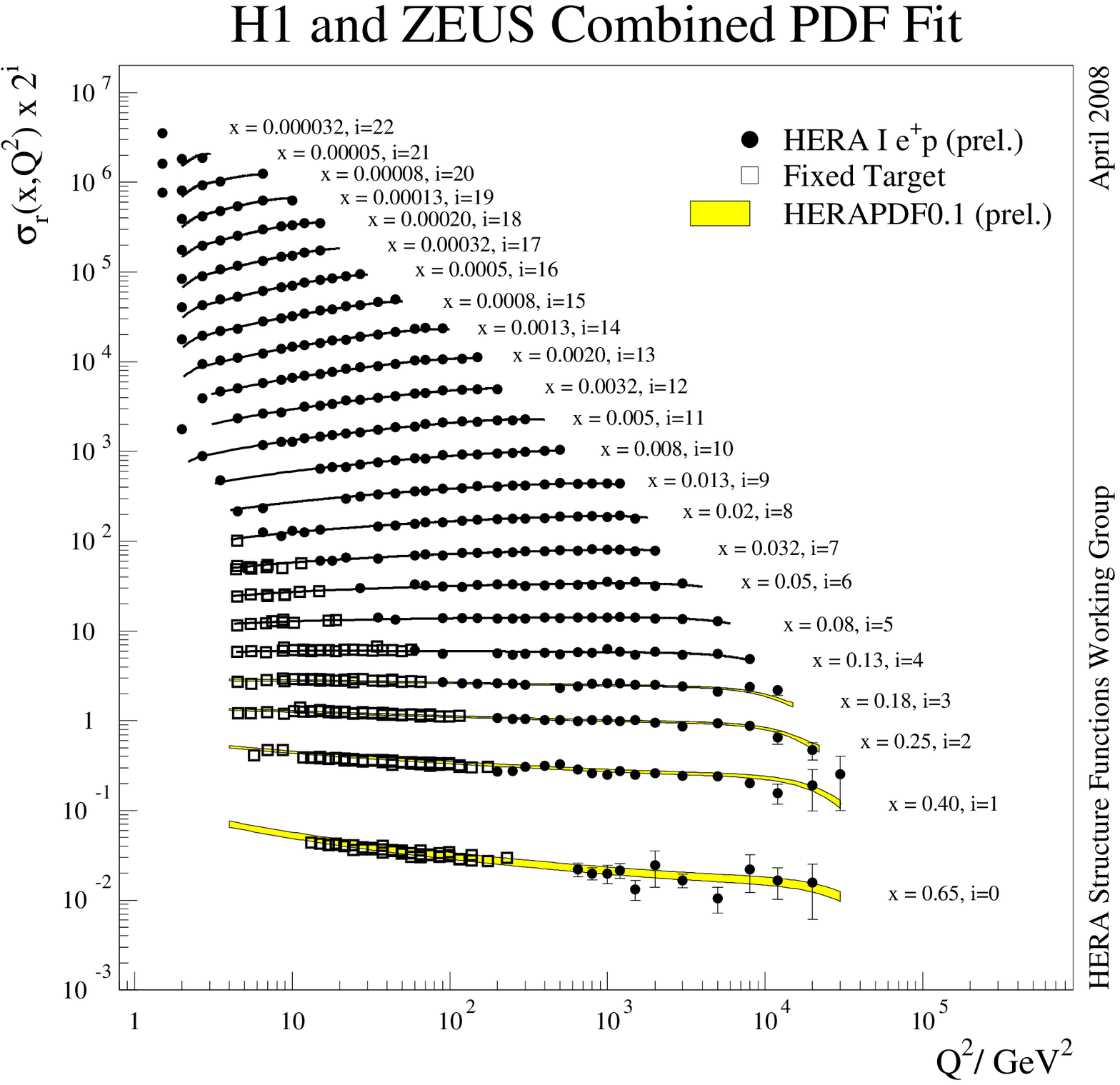}
\includegraphics[width=0.48\textwidth]{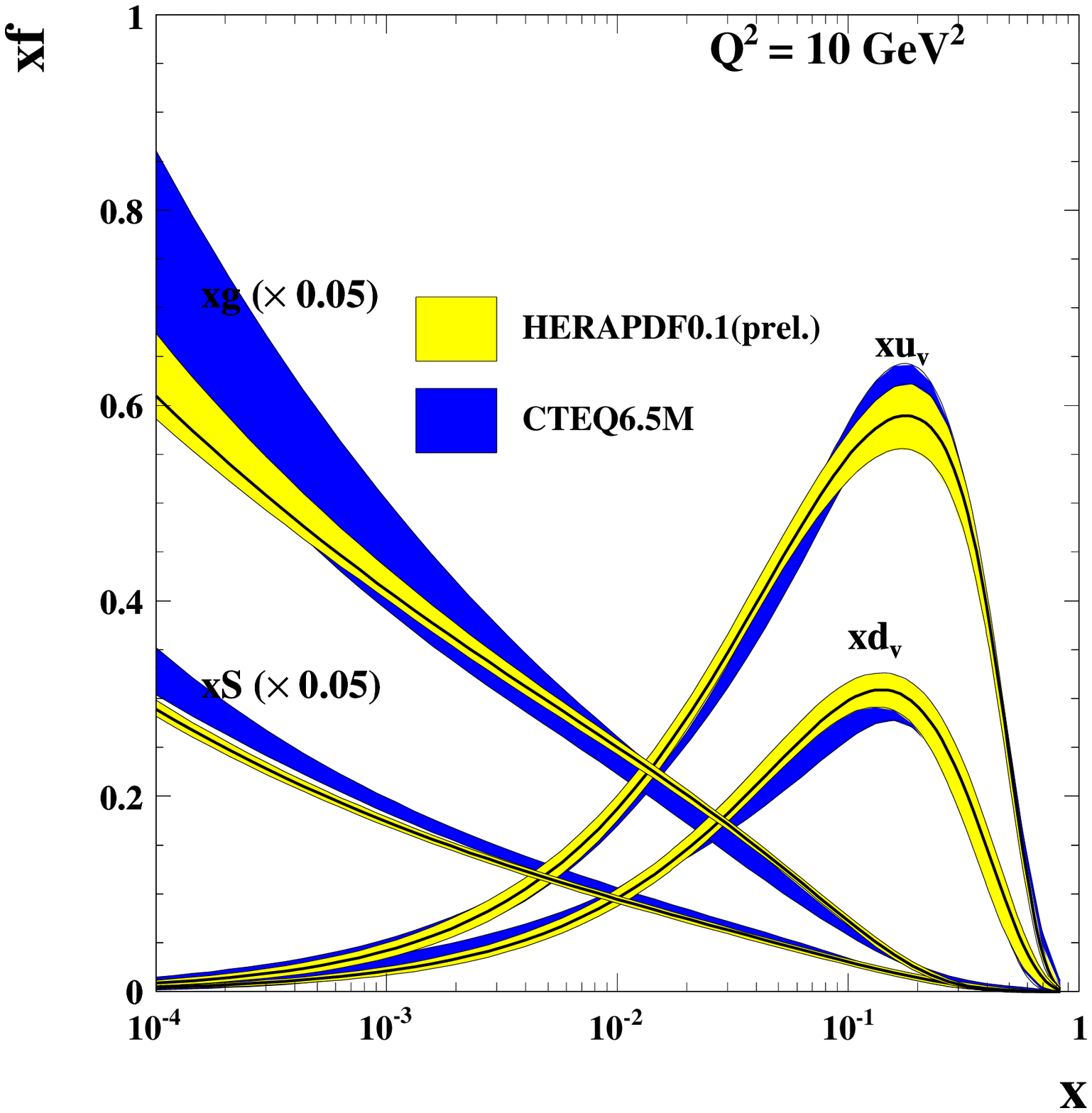}
\caption{LEFT: HERA combined NC reduced cross section $\sigma_r$ as a
  function of $Q^2$ at different $x$ values. The predictions of the
  HERAPDF0.1 fit are superimposed. RIGHT: HERAPDF0.1 PDFs at $Q^2 = 10
  \unit{GeV^2}$ compared to the PDFs from CTEQ6.5M. 
}   
\label{fig:PDFs}  
\end{figure}

The H1 and ZEUS collaborations have both used their data to perform
reduced cross section and $F_2$ measurements as well as NLO QCD PDF fits 
\cite{ref:PDFa} \cite{ref:PDFb}. 
As these measurements are limited by systematic effects, both
data sets have now been combined (Fig~\ref{fig:PDFs}
left) using a 'theory-free' Hessian fit such, that both experiments
'calibrate' each other \cite{h1:zeus:comb}. 
This results in a common data set with reduced
systematic uncertainties, which is then used as input to the
HERAPDF0.1 fit (Fig~\ref{fig:PDFs} right). The new PDFs
\cite{h1:zeus:PDFs} are of impressive precision as compared to the fit
of each dataset alone, as well as to the global parton analyses
\cite{Nadolsky:2008zw} \cite{Martin:2007bv}.  

\section{Neutral and Charged Currents at HERA~II}
\label{sec:NCCChera2}
Exploiting the full luminosity ($\approx 175 \unit{pb}^{-1}$) for a given
lepton-beam charge, and polarization, precise measurements of high-$Q^2$ 
NC and CC cross sections in $e^{-}p$ collisions are performed 
\cite{ref:NCele} \cite{ref:CCele}, allowing for the direct observations 
of the effects of the weak interactions in DIS. The structure function
$x\tilde{F_3}$ is obtained from the difference of $e^+p$ (from
previous data) and $e^-p$ scattering cross sections (Fig.~\ref{fig:FL}
left). All results agree very well with SM predictions and provide
strong constraints at the EW scale (on e.g. $u$ and $d$ quark
couplings to the $Z$ boson).

\section{First direct measurements of the longitudinal structure
  function $F_L$}
\label{sec:FL}
\begin{figure}[h]  
\begin{center} 
\includegraphics[width=0.49\textwidth]{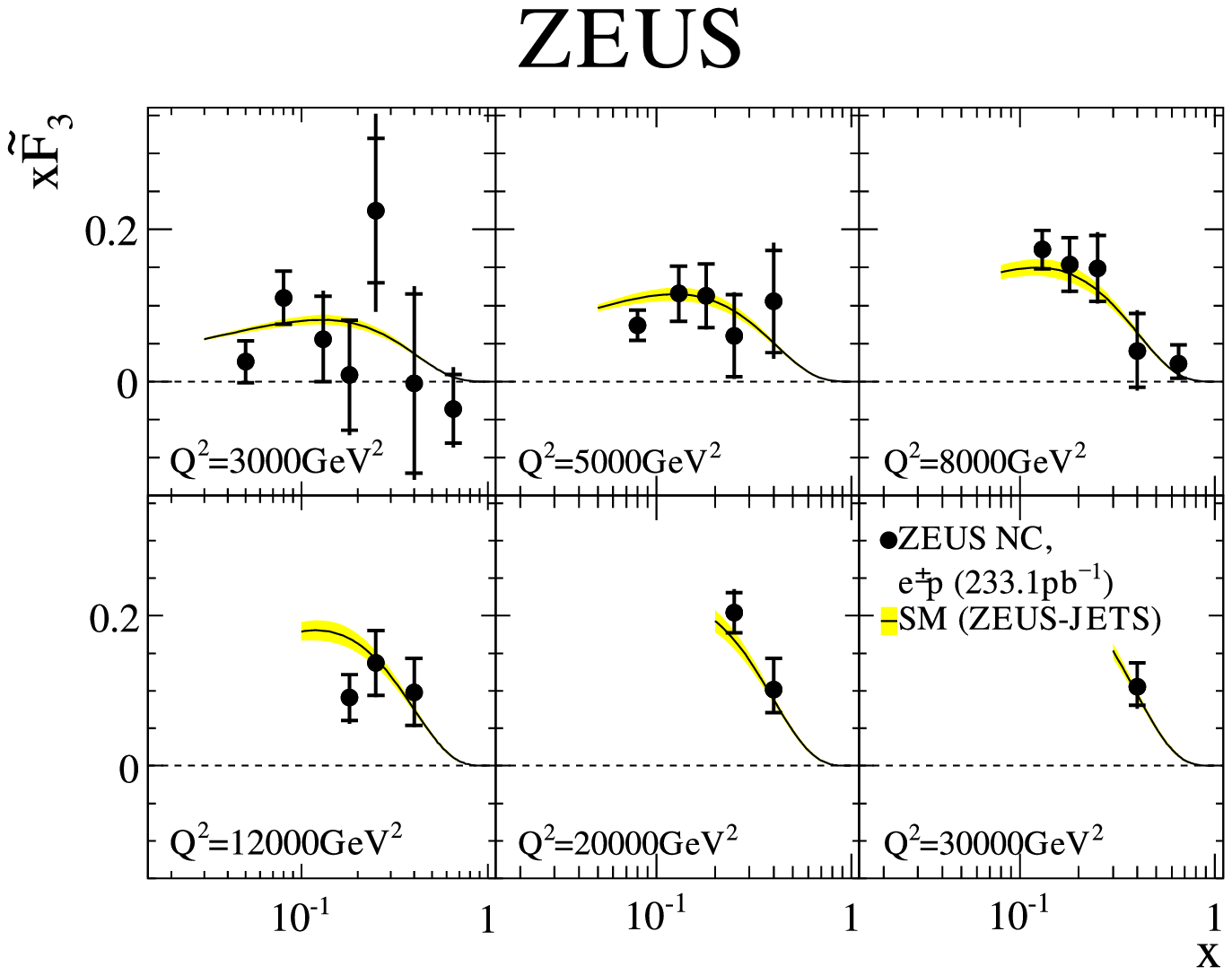}
\includegraphics[width=0.50\textwidth]{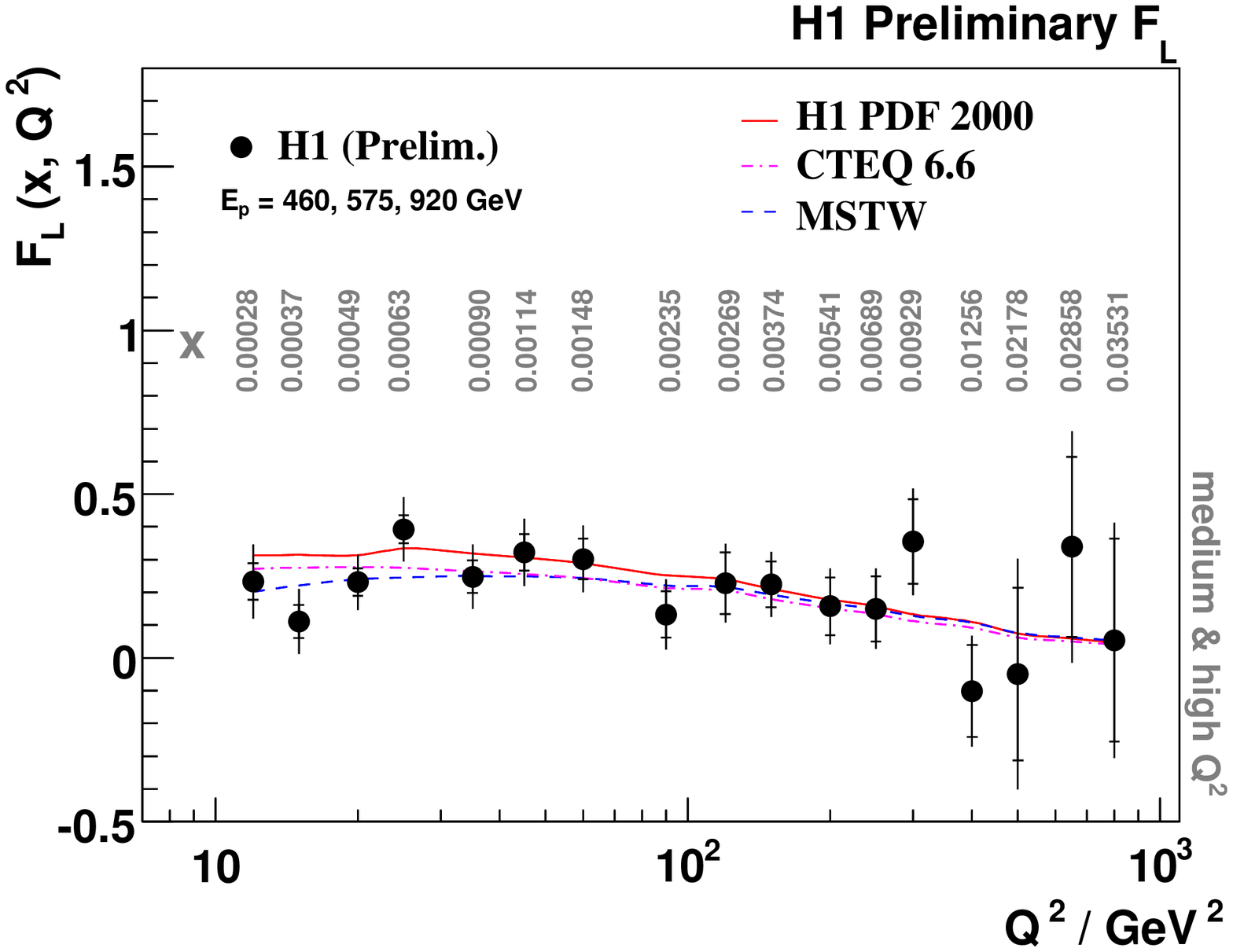}
\end{center} 
\caption{LEFT: The structure function $x\tilde{F_3}$ plotted as a
  function of $x$ in bins of $Q^2$. RIGHT: The $F_L$ structure function
  shown as a function of Q2 at the given values of x.}  
\label{fig:FL}  
\end{figure}

The first direct measurements of ${F_L}$ are performed using
dedicated runs with three proton beam energies (920, 575,
460~\unit{GeV}). 
The longitudinal structure function $F_L(x,Q^2)$ is then extracted at
fixed $(x,Q^2)$ as the slope of 
$\sigma_r$ versus $y^2/Y_+$ \cite{FL:2008tx, FL:ZEUS}.  
The $F_L$ values obtained by H1 are presented in (Fig.~\ref{fig:FL}
right) as a function of $Q^2$. The data are consistent with the SM
predictions evaluated by the MSTW and CTEQ groups, confirming the
applicability of the DGLAP evolution framework at low Bjorken $x$. 

%
%
%

%
\end{document}